\documentstyle[preprint,prc,aps]{revtex} 
\begin{document}
\pagestyle{plain}
\newcount\eLiNe\eLiNe=\inputlineno\advance\eLiNe by -1
\title{
On the atomic resonances in the  $0 \nu 2EC$ transitions}


\author{S. Wycech  and Z. Sujkowski   
\address{ So{\l}tan Institute for Nuclear Studies, Hoza 69, 
Pl-00-681 Warsaw, Poland }}

\maketitle

\begin{abstract}
The nuclear  method to discover  Majorana 
neutrinos  is the neutrinoless double $ \beta $ decay. 
An interesting alternative is offered by the inverse 
process, neutrinoless radiative double electron capture, 
accompanied by  a photon emission. 
Two different  mechanisms seem plausible: the magnetic type 
radiation by an initial electron and  the  
resonant electric type radiation  by the final atom. 
The physical background for these processes is 
calculated. 
\end{abstract}


\section{Introduction}

The search for  neutrino-less double decay,  $ 0\nu\beta \beta$,  
decay is a major challenge of  to-day's neutrino physics.
The process, if found, will prove unambigously 
the  Majorana nature of neutrino. It will also provide a 
sensitive measure of the neutrino mass.  
Nuclear physics, with the studies of $ 0\nu\beta \beta$ 
transitions  is expected to resolve this question for  the electron 
neutrino, $\nu_{e} $. If it is a Majorana particle then  
by definition it  is identical  to its charge conjugate. Thus 
the neutrino  produced in a weak process on one nucleon 
may be absorbed by another nucleon. The outcome is a 
nuclear reaction 
\begin{equation}
\label{1}
 (A,Z-2) \rightarrow (A,Z) +e +e 
\end{equation}  
 Amplitudes for such a process are proportional to the 
Majorana neutrino mass. 
For the description of the problems involved 
we refer to reviews   \cite{BOE87}, \cite{MOH01}, \cite{ELL02}, 
\cite{EJI00}, \cite{KLA01}, \cite{SUJ03}.

The  experimental  check of  transition (\ref{1}) requires detection of 
two correlated  electrons of a given total energy. Such measurements 
are difficult not only because of low rates 
and the dominating random background, but also as a result of the  
high physical background.  The latter is due to the 
neutrino accompanied double beta process $ 2 \nu\beta \beta$.  Some of these
difficulties may be overcome by studying the inverse  process of  neutrino-less double 
electron capture  accompanied by a photon emission \cite{SUJ03}, \cite{SUJ02}. There are
 several experimental advantages. 
The monoenergetic  photon provides a convenient experimental signature. 
Other advantages include the favourable ratio of the  $0 \nu 2EC$ 
to the competing  $2\nu 2EC$ capture rates as opposed to that 
of  $0 \nu \beta \beta $ to  $ 2\nu \beta \beta $. 
This point is discussed  in  detail below.  
Another  important advantage of the  $0 \nu 2EC$ process 
is the existence of the coincidence trigger to suppress 
the random background.
These advantages offset, in part, the longer lifetimes 
for the  $0 \nu 2EC$ decays.

Chances for the capture process have been calculated 
in ref.\cite{SUJ02}. The seemingly most likely  process,  
which involves  
two $1S$ electrons and leads to two $1S$ holes in the 
final atom,  does not conserve spin. Allowed is the  capture 
of  $1S,2S$  electron pair with the photon emitted by one of these  
electrons in  intermediate states. The rate increases 
roughly with $ Z^7$, high Z atoms are thus required. 
Still, the  experiments would be  difficult. There is yet 
another process which may be more likely at small energy 
release. This involves  the  virtual captures of $1S,1S$ 
electrons and the radiation process in the final excited atom. 
A resonance enhancement of the capture rates is  
predicted  \cite{WIN55}, \cite{BER83}, \cite{ZSU03}, 
when the energy release  $Q$ is 
comparable to the $2P-1S$ atomic level difference.  
Away from the resonance the rates depend only slowly on 
$Q$, in strong contrast with the $0 \nu \beta \beta $ 
decays. This makes studies of decays to excited states in
final nuclei feasible, thus enhancing chances of locating the 
resonances. Candidates for such studies have been considered 
\cite{BER83}, \cite{SUJ02}, and the experimental feasibility 
is found encouraging,\cite{ZSU03}.

This paper  consists of two sections. Section II  gives a 
brief presentation of two basic  radiation  mechanisms that 
occur in the  $0 \nu 2EC$ transitions. 
In Section III the physical background for these two 
modes is  calculated.

\section{Two modes of radiative $0 \nu 2EC$ transitions }

The rate $\Gamma (0 \nu \beta \beta )$ for the  
neutrino-less double beta decay  may be factorised 
into nuclear and leptonic parts  (see e.g. \cite{EJI00},\cite{DOI85}, \cite{SUH98})
\begin{equation}
\label{a1}
 \Gamma (0 \nu \beta \beta )= 
G^{0\nu} \mid M^{0\nu} \mid ^2 (m_{\nu}/m_{e})^2.
\end{equation}  
This equation  involves nuclear matrix elements 
$M^{0\nu}$, the leptonic contributions including 
 final state electron wave functions and  final phase space 
elements are contained in $ G^{0\nu}$. The neutrino mass 
factor $ m_{\nu}$ reflects the chance for the left handed 
neutrino emitted from one nucleon to 
have the righ-handed component required for the subsequent 
absorption on another nucleon. 
The reverse process of  double electron capture  
\begin{equation}
\label{a2}
 (A,Z) + e + e  \rightarrow (A,Z-2) 
\end{equation}  
is not allowed by  the energy-momentum conservation,  
emission of a third body is necessary. 
In the following we consider the photon emission to 
fulfill this requirement: 
\begin{equation}
\label{a3}
 (A,Z) +e +e   \rightarrow (A,Z-2) +  \gamma 
\end{equation}  
Again, the capture rate may be factorised 
into the nuclear, leptonic and photonic  terms  
\begin{equation}
\label{a4}
 \Gamma (0 \nu \gamma )= 2\pi
\int \frac{ d{\bf  q}}{(2\pi)^3} \delta(Q-q )
G^{2EC} [ \frac{M^{0\nu} m_{\nu}}{4\pi}]^2
  \mid M^{\gamma}(q) \mid^2 
\end{equation}  
where $Q$ is the photon energy given by the mass difference of 
the initial and final atoms reduced by the energies of 
two electron  holes left in the final state.
The factors involved in eq.(\ref{a4}) differ from 
those of eq.(\ref{a1}) 
by the final phase space, the transition energy and  electron  
wave functions. The nuclear transition element is of similar nature 
though it connects different nuclei. It is given  by the 
terms in brackets   which describe the propagation 
of neutrino between two nucleons. These contain the  neutrino 
mass and  the nuclear matrix element of  "neutrino exchange 
potential". For the dominant  Gamow-Teller transitions   the latter 
is 
$M^{0\nu} = 
<Z-2 \mid \frac{{\bf \sigma }{\bf \sigma'}exp(-r q_{\nu})}{r} \mid Z> $ 
where $r$ is the distance between the two nucleons and $q_{\nu} $  
is the average energy needed to generate neutrino of very short 
propagation range. Calculations yield $M^{0\nu} \approx 1/fm$,
 \cite{EJI00},\cite{SUH98}.    

An additional factor $M^{\gamma}$ introduced into eq.(\ref{a4}) 
denotes  the probability of photon emission. Two cases are likely 
to offer a good chance for the  experimental detection.  The first 
one 
occurs when an electron from $2S$ or $1S$ state radiates before it  
is captured by the nucleus. 
In this case, 
\begin{equation}
\label{a5}
\mid M^{\gamma}(q) \mid^2 =  \frac{e^2}{2q m_e^2} f_M 
\end{equation}  
where $e$ is the electron charge and $\sqrt{2q}$ is 
the  photon wave  normalisation. The first term  gives  
the  order of magnitude 
estimate. Finer calculations based   
on Glauber-Martin \cite{GLA56},\cite{MAR58}, theory for 
the electron radiating in the Coulomb field 
give the corrective  factor $f_{M}\approx 1$ for the magnetic 
transition that takes place in these circumstances, 
\cite{SUJ02},\cite{ZSU03}.

The second mode  of radiative transitions comes from a 
different scenario, indicated in  figure 1. 
Two  $1S$ electrons may be  captured in a virtual process which 
generates a final atom with two $1S$ electron holes. 
This final atom radiates and one of the holes is filled. 
That is a resonant-like situation and close to the resonance 
\begin{equation}
\label{a6}
\mid M^{\gamma}(q) \mid^2  =  
\frac {\Gamma^r (2P \rightarrow 1S) }{ [ q - Q_{res}]^2+ [ \Gamma^r/2 ]^2}
\frac {\pi}{q^2}
\end{equation}  
where $\Gamma^r $ is the radiative width of the final 
two-electron-hole atom.  There are a  number of resonant 
situations. The most important one happens with  
$  Q_{res} =  E(2P)-E(1S) $  
that is when the $0 \nu 2EC$ transition energy $Q$ coincides with the 
$2P-1S$ electric radiative transition in the final nucleus 
(the $K_{\alpha}$ transition indicated in eq.\ref{a6}).
These resonant situations may greatly enhance the rate.
Practical interest requires $ \mid  Q_{res} -Q \mid < 1$ KeV. 
There are several targets likely to fulfill this condition, 
\cite{BER83},\cite{SUJ03}. 

\begin{figure}[hp]
\begin{center}
\vspace{7.0cm}
\includegraphics{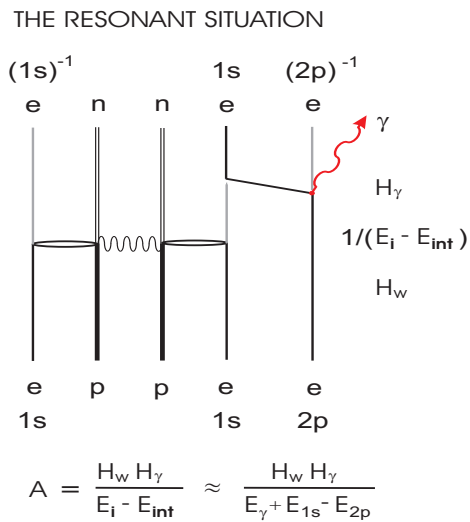}
\caption{The diagram for the $ 0 \nu 2EC$,   
double electron capture.  Indicated is the 
resonance situation that occurs in the intermediate state:
after the  capture - before the  radiation process. 
 $H_W ,H_{\gamma}$ are the weak and radiative hamiltonians 
respectively, $E_i ,E_{int}$ are energies of the  
initial and  intermediate states, $E_{1s},E_{2p}$ are negative 
atomic state energies.  } 
\label{fig:resonance}
\end{center}
\end{figure}

\section{The physical   background}
      
Assuming the photon energy resolution to be $D$ 
we calculate now the ratio of the signal from the $0\nu \gamma $
to the physical background due to the $\nu \nu \gamma $ 
transitions in the double electron capture. This ratio is defined as   
\begin{equation}
\label{b1}
R_{s/b} = \frac{\Gamma (0 \nu \gamma )}{\Gamma (\nu \nu \gamma )N_D}
\end{equation}  
where $N_D$ is the fraction of photons from the dominant 
$\nu \nu \gamma $ decay mode emitted into the region  
from the end of the spectrum $Q$ down to $Q-D/2$.
For the two neutrino process the radiative  rate is
\begin{equation} 
\label{b2}
 \Gamma (\nu \nu \gamma )= 
\int \frac{2\pi d{\bf  q}d{\bf  p}d{\bf  p'}}{(2\pi)^9 2E(p)2E(p')}
 \delta(Q -E(p)-E(p')-q) 
G^{2 EC } \mid M^{2\nu} \cdot M^{\gamma}(q) \mid^2,  
\end{equation}  
where the nuclear matrix element  $ M^{2\nu}$ differs 
from  $ M^{0\nu}$. For  the  Gamow-Teller transitions one has  
$ M^{2\nu} = <Z-2 \mid {\bf \sigma }{\bf \sigma' } \mid Z> $, and
calculations, \cite{SUH98}, \cite{EJI00}, 
yield $ M^{2\nu} \approx 1 $.
The matrix elements $ M^{2\nu}$ for 
the neutrino and $ M^{0\nu}$ for the 
neutrino-less decays are almost  energy independent. 
The leptonic factor $ G^{2 EC }$  is also constant 
as opposed to the equivalent factor in the double 
$\beta $ decays. The required ratio $R_{s/b} $  
is thus given by the phase space. 
Let us present eq.(\ref{b2}) as an integral over the photon 
energy distribution $ W(q)$: 
\begin{equation}
\label{b3}
 \Gamma (\nu \nu \gamma)= \int_{0}^{Q} W( q ) dq. 
\end{equation}
The background contribution follows as  
\begin{equation}
\label{b4}
\Gamma (\nu \nu \gamma )N_D = \int_{Q-D/2}^{Q} W( q ) dq. 
\end{equation}
>From eq.(\ref{b2}) one obtains 
\begin{equation}
\label{b5}
W( q )= 
\frac{G^{2 EC } \mid M^{2\nu}\mid ^2 }{(2\pi)^5 3}
 q^2( Q-q)^3  \mid M^{\gamma}(q)\mid^2. 
\end{equation}

First we consider the magnetic type transition  related to 
the $1S,2S$ electron capture. In this case $M^{\gamma}$ 
given by eq.(\ref{a4}) is constant  and  the ratio of total 
rates becomes :
\begin{equation}
\label{b6}
R_{0\nu/2\nu} = \frac{\Gamma (0 \nu \gamma )}{\Gamma (\nu \nu \gamma )}=
\frac{ 120 \pi^2 m_{\nu}^2\mid M^{0\nu}\mid ^2} {Q^4 \mid M^{2\nu}\mid ^2}
\end{equation}  
which  gives  the relative frequency of  the non-neutrino 
 to the   corresponding  two-neutrino radiative transitions.
For characteristic values  $ Q= 1$ MeV and $  m_{\nu} =1 $ eV one
obtains $ R_{0\nu/2\nu} = 5 \cdot 10^{-5}$. The contribution 
from the physical background to the region 
of the monoenergetic photon  depends on the energy resolution: 
\begin{equation}
\label{b7}
R_{s/b} = 
\frac{3 \cdot  2^7  \pi^2 m_{\nu}^2\mid M^{0\nu}\mid ^2} {D^4 \mid M^{0\nu}\mid ^2}
\end{equation}  
For 
$ Q= 1$ MeV, $  m_{\nu}=1 $ eV and $ D= 3$ KeV one
obtains very low background  $ R_{s/b} = 2 \cdot 10^{6}$. 

Now we turn to the case of  the resonant electron captures. 
The rate of the signal to the background depends strongly 
on the relation of the transition energy to the resonance energy. 
Assume that the situation of a  nearby resonance is  
materialised with  $ Q > Q_{res} $,  that is the resonance 
transition is located 
within the photon spectrum. Also,  we assume the line to be within 
the photon energy resolution band $ \mid Q - E_{res}\mid < D/2 $. 
We exclude for a moment the optimal but highly unlikely situation 
of $ \mid~Q~-~E_{res}\mid~<~\Gamma^r $ corresponding to  the line 
coinciding with the end of the photon spectrum.  
With these limitations one obtains  
\begin{equation}
\label{b8}
R_{s/b} \approx 
\frac{6 \pi  m_{\nu}^2 \mid M^{0\nu}\mid ^2}
{ [ Q- Q_{res}]^3  \mid M^{2\nu}\mid ^2}
\frac {\Gamma^r (2P \rightarrow 1S) }{ [ Q- Q_{res}]^2+ [ \Gamma^r/2 ]^2}.
\end{equation}  
Beyond  the range of  natural widths  ($0.05- 0.10$ KeV)  
the ratio $R_{s/b}$ falls down as  $[ Q- Q_{res}]^{-5}$.   
 For $ Q- Q_{res} = 1 $ keV  
the signal to background rate is still very convenient 
$ R_{s/b} \approx  10^{4}$.  However, the conditions deteriorate 
quickly with the increasing separation. 

Formula (\ref{b8}) is valid for 
$  Q- Q_{res}> \Gamma^r  $.    
The other two situations, i.e. that of $ Q < Q_{res} $  
when the centre of the 
atomic line is beyond the photon  spectrum, or that of a 
"direct hit"  into the line,   $ \mid Q - E_{res}\mid < \Gamma^r$,  
offer the luxurious ratio  $ R_{s/b} \approx  10^{9}$, 
which means  no physical background.

Support from the KBN project 5P 03B 04521  is acknowledged.

\end{document}